\documentclass[a4paper,12pt]{article}
\usepackage{anysize}
\marginsize{2cm}{2cm}{1cm}{1cm}
\usepackage{amssymb}
\usepackage{amsmath}
\usepackage{natbib}
\usepackage[Large]{subfigure}
\usepackage{graphicx}
\usepackage{lscape}
\usepackage{float}
\usepackage[title,titletoc,toc]{appendix}
\restylefloat{table}
\newtheorem{theorem}{Theorem}

\newtheorem{proposition}[theorem]{Proposition}
\newtheorem{remark}[theorem]{Remark}

\begin{document}
\newcommand{\sgn}{\mathop{\mathrm{sgn}}}

\title{Parametric Risk Parity }
\author{Lorenzo MERCURI and Edit RROJI}
%
%
\maketitle
\abstract{Any optimization algorithm based on the risk parity approach requires the formulation of portfolio total risk in terms of marginal contributions. In this paper we use the independence of the underlying factors in the market to derive the centered moments required in the risk decomposition process when the modified versions of Value at Risk and Expected Shortfall are considered.  \newline\indent
The choice of the Mixed Tempered Stable distribution seems adequate for fitting skewed and heavy tailed distributions. The ensuing detailed description of the optimization procedure is due to the existence of analytical higher order moments. Better results are  achieved in terms of out of sample performance and greater diversification.
}
\section{Introduction}

Nowadays there is much more emphasis on the sources of risk rather than just only on the levels. In addition to the interest on the marginal contribution to risk of a particular factor, we have to deal with new concepts  such as the Risk Parity. It is an approach in portfolio management which focuses on allocation of risk rather than on the capital allocation \citep[see][for further details] {Denis2011} it suggests that in a well-diversified portfolio all asset classes should have the same marginal contribution to the total risk of the portfolio.
 
In financial literature non parametric methods based on historical simulation have been studied deeply but, as observed in \cite{meucci2009risk}, an approach that takes into account only  past realizations of the variables of interest depends on the choice of the time interval. Stability issues for estimates  require large sample sizes (see for example \cite{martellini2010improved}, \cite{hitaj2013portfolio} in the context of sample moments applied to the portfolio selection problem) but on the other hand realizations observed in the farther past can be less realistic since market conditions may have changed meanwhile. A simple answer to this problem is the use of exponentially decaying weights for the observations, i.e instead of giving equal weight to each observation in the past we consider as more relevant recent realizations. But in so doing we may wrongly give little importance to scenarios in the past that realized in similar conditions to the today's market. Parametric distributions enough flexible to fit time series of financial returns can be a starting point for procedures based on estimates of moments that present statistical properties even for not so large  sample sizes.

Recently a new class of distributions, named Mixed Tempered Stable distribution (MixedTS hereafter), has been introduced in \cite{Rroji2014, Rroji2014Mixed}. The idea behind is to generalize the Normal Variance Mean Mixtures (NVMM henceforth) substituting the normality assumption  with the Tempered Stable distribution \citep[see]{BookContTankov}. In this way the new distribution overcomes some limits of the NVMM. In particular the MixedTS is more flexible in capturing the higher moments since in the NVMM the sign of skewness is given by the sign of the drift parameters and the level depends on the mixing random variable and drift parameter while in the MixedTS, the asymmetry depends also on the tempering parameters of the Tempered Stable distribution.\newline A similar argument holds for the kurtosis since for particular choice of the tempering parameters, the tail behavior of the MixedTS varies from semi-heavy (i.e. the tail decays exponentially) to heavy (power law decay), while the tail behavior for the NVMM depends only on the tail behavior of the mixing random variable \citep[see][for a complete discussion on tail behavior]{barndorff1982normal}. Here, we find an advantage of the MixedTS in modeling
 financial returns since we do not need to know a priori if we have to consider a heavy or semi-heavy distribution.

The main contribution of this paper is the introduction of a general setup for obtaining risk parity portfolios by modeling directly the underlying  factors in a given market. For factor identification, we apply the Independent Component Analysis (ICA) introduced
in \cite{art:Comon1994}. Details and algorithms on this subject are given in  \cite{BOOK:hyvarinen2001}. 
Exploiting the ability of  ICA to decompose observed signals in independent random variables, in the proposed approach we need only to model each individual component since the dependence structure of factors is captured by the mixing matrix obtained through the algorithm.  

From the Euler  theorem for homogeneous functions we have that a  homogenous risk measure can be written as a weighted sum of the marginal risk contribution where the weights are the exposures to the factors \citep[see][for a complete treatment]{Tasche99}. Consequently, risk parity portfolios are achieved as a solution of a constrained minimization problem as proposed for example in \cite{maillard2008properties}.\newline  
In this paper we focus on three standard homogeneous risk measures: Volatility, Value at Risk (VaR) and Expected shortfall (ES). In particular for the last two measures we consider the modified versions proposed in \cite{Zangari1996} for the VaR and in \cite{boudt2007estimation} for the ES. The idea behind both modified measures is to consider asymptotic expansions for the underlying distribution based on the first four moments that, in our approach, can be easily derived using the ICA approach and assuming factors to be MixedTS-distributed.


The outline of the paper is as follows. In Section 2 we briefly recall the risk parity approach and its connection with other portfolio optimization methods. The main results concerning the Mixed Tempered Stable distribution are reviewed in Section 3 while in Section 4 we analyze the risk parity approach for portfolio optimization using the modified VaR and the modified ES. Empirical results are given in Section  5 and  Section 6 concludes the paper.    
\section{Portfolio construction using the Risk Parity approach}

Risk parity is an approach of allocating risk rather than capital. It overcomes some of the limits of standard approaches like for example  mean-variance optimization. Indeed, as observed in \cite{maillard2008properties}, the mean-variance approach has two drawbacks in practice. First, optimal portfolios seem to be  concentrated in a few assets. Second, small changes in the estimated parameters give rise to relevant modifications in the optimal portfolio that as remarked by \cite{merton1980} is more relevant in the case of portfolio expected return estimation. To avoid this lack of stability, researchers proposed several regularization techniques. The most used are resampling of the objective function proposed by \cite{michaud1989markowitz} and shrinkage estimators of the covariance matrix developed in \cite{ledoit2003improved}. In literature we also find heuristic approaches that do not require return estimation like Equally weighted (EW), Equal risk contributions (ERC) or Minimum Variance (MV) portfolios. Through these methods, we put constraints directly  on portfolio weights and do not require advanced programming issues. These methods are not completely distant from each other. For example Equally weighted portfolios can be seen as a particular case of Equal risk contributions supposed we have the same risk and the same correlation for all the factors.
\newline
A common way of expressing portfolio returns is as a linear combination of factor returns ($F$) with weights given by the portfolio exposures in $\beta$:
\begin{equation}
r=\beta' F
\label{Fact:Mod}
\end{equation}Identifying all factors that influence portfolio returns is not an easy task but once we get them a very important concept when dealing with risk analysis is the marginal contribution to risk (MRC) of a factor or an asset class defined as:
\begin{equation}
MRC_{i}=\frac{\partial R(r)}{\partial \beta_{i}}
\label{Marg:Risk}
\end{equation}This quantity represents the additional risk of our portfolio  for each additional unit of exposure to the i-th factor.  Of particular interest is the product of the exposure with the marginal contribution to risk known as total risk contribution (TRC):
\begin{equation}
TRC_{i}=\beta_{i}\frac{\partial R(r)}{\partial \beta_{i}}
\label{totalriskcontrib}
\end{equation}The use of TRC makes risk attribution easier to understand as it becomes the split of risk in portions that are additive and constitute the portfolio total risk.\newline
Risk parity, as other portfolio optimization rules, aims at identifying portfolio weights (or exposures) that satisfy a certain criteria. In practice, TRC must be the same for each factor considered in the portfolio construction. \cite{maillard2008properties} propose to perform the following minimization to get the desired weights:

\begin{equation}
\begin{aligned}	
& \underset{\beta}{\text{minimize}}
& & \mathrm \sum_{i=1}^{N} \sum_{j=1}^{N} (TRC_{i}-TRC_{j})^2 \\
& \text{subject to}
& & \sum_{i=1}^{N}\beta_{i} = 1,  \\
&&& 0 \leq \beta_i \leq 1 \;  i = 1, \ldots, N.
\end{aligned}
\label{Risk:ParProb}
\end{equation}where the inequality constraints refer to the no-short selling conditions.\newline
It is worth noting that the objective function in the optimization problem introduces a penalty when TRCs are different from each other. In this way, the resulting portfolio has similar TRC for each considered factor. 

\section{Mixed Tempered Stable distribution}
In this section we review the main results on the Mixed Tempered Stable introduced in \cite{Rroji2014} and investigate the methods for computing risk measures in  the univariate case. Before introducing the MixedTS we start from the definition of Normal Variance Mean Mixtures. \newline
NVMM models are based on the normality assumption while we try to generalize this concept. In fact a Normal Variance Mean Mixture has the form :
\begin{equation}
Y= \mu_{0}+\mu V + \sigma \sqrt{V}Z
\label{defnvmm}
\end{equation}
where the parameters $\mu_{0},\mu  \in \Re$ and $Z \sim N(0,1)$. $V$ is continuously  distributed on the positive half-axis. The main idea behind the MixedTS is to substitute the normality assumption for the r.v. $Z$ in formula (\ref{defnvmm}) with the Tempered Stable that ensures more flexibility to the new distribution.\newline
We recall that the Tempered Stable distribution is obtained by multiplying the L\'evy density of an $\alpha$-Stable with a decreasing tempering function (\cite{BookContTankov}). Tail behavior changes from heavy to semi-heavy characterized by exponential instead of power decay and the existence of the conventional moments is ensured.
 \cite{Twd1984} introduced the one side Tempered Stable distribution by exponentially tilting the tail of a positive Stable distribution. \cite{Rosinski} generalized the tempered stable distributions and classified them according to their L\'evy measure. With this generalization it is also possible to obtain distributions with the whole real axis as support. \cite{KT2013} observed that the Tempered Stable defined on real axis can be obtained as a difference of two independent one sided Tempered Stable. This distribution and the corresponding process has been widely applied in finance \citep[see][for modeling asset returns and the recent textbook \cite{rachev2011financial}]{kuchler2014exponential,mercuri2008option}.

In this paper we consider a parametric distribution, the Mixed Tempered Stable, for modeling asset returns and use it in risk computation. We say that a continuous random variable $\mathbf{Y}$ follows a Mixed Tempered Stable distribution if:
\begin{equation}
Y\stackrel{d}{=}\mu_{0}+\mu V+\sqrt{V}X
\label{Def:MixTempStab}
\end{equation} where $X\left|V\right.\sim stdCTS(\alpha,\lambda_{+}\sqrt{V},\lambda_{-}\sqrt{V})$ is Standardized Classical Tempered Stable distributed ($stdCTS$ \cite{KT2013}). V is an infinitely divisible distribution defined on positive axis and its m.g.f always exists. The logarithm of the m.g.f. is:
\begin{equation}
\Phi_{V}(u)=\ln\left[E\left[\exp\left(uV\right)\right]\right]
\label{Def:CharactSub}
\end{equation}We compute the characteristic function for the new distribution and apply the law of iterated expectation:
\begin{eqnarray}
E\left[e^{iuY}\right]&=&E\left[E\left[\left.e^{iu\left(\mu_{0}+\mu V +\sqrt{V}X\right)}\right|V\right]\right]\nonumber\\
&=&e^{i u \mu_{0}}E\left[e^{\left[u\mu + L_{stdCTS}\left(u;\ \alpha, \ \lambda_{+}, \ \lambda_{-}\right)\right]V}\right]\nonumber\\
&=&e^{i u \mu_{0}+\Phi_{V}\left(u\mu + L_{stdCTS}\left(u;\ \alpha, \ \lambda_{+}, \ \lambda_{-}\right)\right)
}\nonumber\\
\label{FinalChar}
\end{eqnarray}The characteristic function identifies the distribution at time one of a time changed L\'{e}vy process and the distribution is infinitely divisible. Despite the fact that this distribution has nice features from a theoretical point of view, it allows a dependence of the standard higher moments not only on the mixing r.v but also on the Standardized Classical Tempered Stable distribution parameters . As observed in \cite{Rroji2013}, it is important to have a flexible distribution for accommodating the differences in terms of asymmetry and tail heaviness. \newline
\begin{proposition}
The first four moments of the MixedTS have an analytic expression since:
\begin{equation}
\left\{ 
\begin{array}{l}
E\left[Y\right]=\mu_{0}+\mu E\left[V \right] \\
Var\left[Y\right]=\mu^2 Var(V)+ E\left[V \right]\\
m_{3}\left(Y\right)=\mu^3 m_{3}\left(V\right)+3\mu Var(V)+\left(2-\alpha\right)\frac{\left(\lambda_{+}^{\alpha-3}-\lambda_{-}^{\alpha-3}\right)}{\left(\lambda_{+}^{\alpha-2}+\lambda_{-}^{\alpha-2}\right)} E\left[V\right]\\
m_{4}\left(Y\right) =\mu^4 m_{4}(V)+6 \mu^2 E\left[ \left(V-E(V)\right)^2 V \right]+4 \mu \left(2-\alpha\right)\frac{\lambda_{+}^{\alpha-3}-\lambda_{-}^{\alpha-3}}{\lambda_{+}^{\alpha-2}+\lambda_{-}^{\alpha-2}}Var(V)\\
+ (3-\alpha)(2-\alpha)\frac{\left(\lambda_{+}^{\alpha-4}+\lambda_{-}^{\alpha-4}\right)}{\left(\lambda_{+}^{\alpha-2}+\lambda_{-}^{\alpha-2}\right)}E\left[V\right]\\
\end{array}
\right.
\label{Mom:MixedTS}
\end{equation}
Where $m_{3}()$ and $m_{4}()$ are the third and fourth central moments respectively. 
\end{proposition}
See Appendix \ref{DerivMoment} for details on the derivation of the moments. \newline
The choice of using this distribution comes from the fact that if we assume that $V \sim \Gamma(a,\sigma^2)$, we have as special cases some well-known distributions in modeling financial returns. We get the Variance Gamma \citep{MadanSeneta1990, Loregian2012} for $\alpha=2$ and the  Standardized Classical Tempered Stable when $\sigma=\frac{1}{\sqrt{a}}$ and $a$ goes to infinity.\newline
Under the assumption that $V \sim \Gamma(a,\sigma^2)$
\begin{eqnarray*}
E\left[V\right]&=&a \sigma^2\\
Var\left[V\right]&=& a \sigma^4\\
E\left[ \left(V-E(V)\right)^2 V \right]&=&E\left[ \left(V-E(V)\right)^3\right]+E(V)Var(V)= \frac{2}{\sqrt{a}}a^{3/2} \sigma^6+a^2\sigma^6\\
E\left[ \left(V-E(V)\right)^3\right]&=& \frac{2}{\sqrt{a}}a^{3/2} \sigma^6\\
E\left[ \left(V-E(V)\right)^4\right]&=& \left(3+ \frac{6}{a}\right)a^2 \sigma^8\\
\end{eqnarray*}

\begin{remark}
From the scale property of the Gamma r.v. we have that
\begin{equation*}
V\stackrel{d}{=}\sigma^2 \tilde{V}
\end{equation*}where $\tilde{V} \sim \Gamma(a,1)$ and the definition in \eqref{Def:MixTempStab} can be written as:
\begin{equation*}
Y\stackrel{d}{=}\mu_{0}+\tilde{\mu} \tilde{V}+\sigma\sqrt{\tilde{V}}\tilde{X}
\end{equation*}where $\tilde{\mu}=\mu\sigma^2$ and $\tilde{X}\sim stdCTS(\alpha,\lambda_{+}\sigma \sqrt{V},\lambda_{-}\sigma \sqrt{V})$. Note that in this formulation the MixedTS distribution has the same structure of the NVMM defined in \eqref{defnvmm}.
\end{remark}
For univariate random variables, risk measures can be computed directly once we have the characteristic function $\phi_{Y}\left(t\right)$ of a r.v $Y$ since we evaluate its distribution function $F_{Y}\left(y\right)$ using the formula based on the Inverse Fourier Transform:
\begin{equation*}
F_{Y}\left(y\right)=\frac{1}{2}-\frac{1}{2\pi}\int_{-\infty}^{+\infty}\frac{\left[e^{-ity}\phi_{Y}\left(t\right)\right]}{it}\mbox{d}t
\end{equation*}The Value at Risk at he confidence level $\alpha$ is obtained inverting the distribution function:
\begin{equation*}
Var_{\alpha}(Y)=-F_{Y}(\alpha)
\end{equation*} Under the assumption of existence for the $E\left(Y\right)$, the Expected Shortfall is computed using the formula: 
\begin{equation*}
ES_{\alpha}(Y)=E\left[Y\right|\left. Y\leq y_{\alpha}\right]=y_{\alpha}-\frac{1}{\alpha}\int_{-\infty}^{y_{\alpha}}F\left(u\right) \mbox{d}u
\end{equation*}In a multivariate context the distribution function can not obtained trivially since it is based on a model that captures the dependence of assets and requires the computation of multiple integrals. In the next section we present a methodology for computing the portfolio risk measures where the dependence structure of its assets is reconstructed through an ICA analysis and each signal is modeled through the MixedTS ditribution.

\section{Parametric risk decomposition}

\begin{figure}[H]
\centering
\includegraphics[width=1\textwidth]{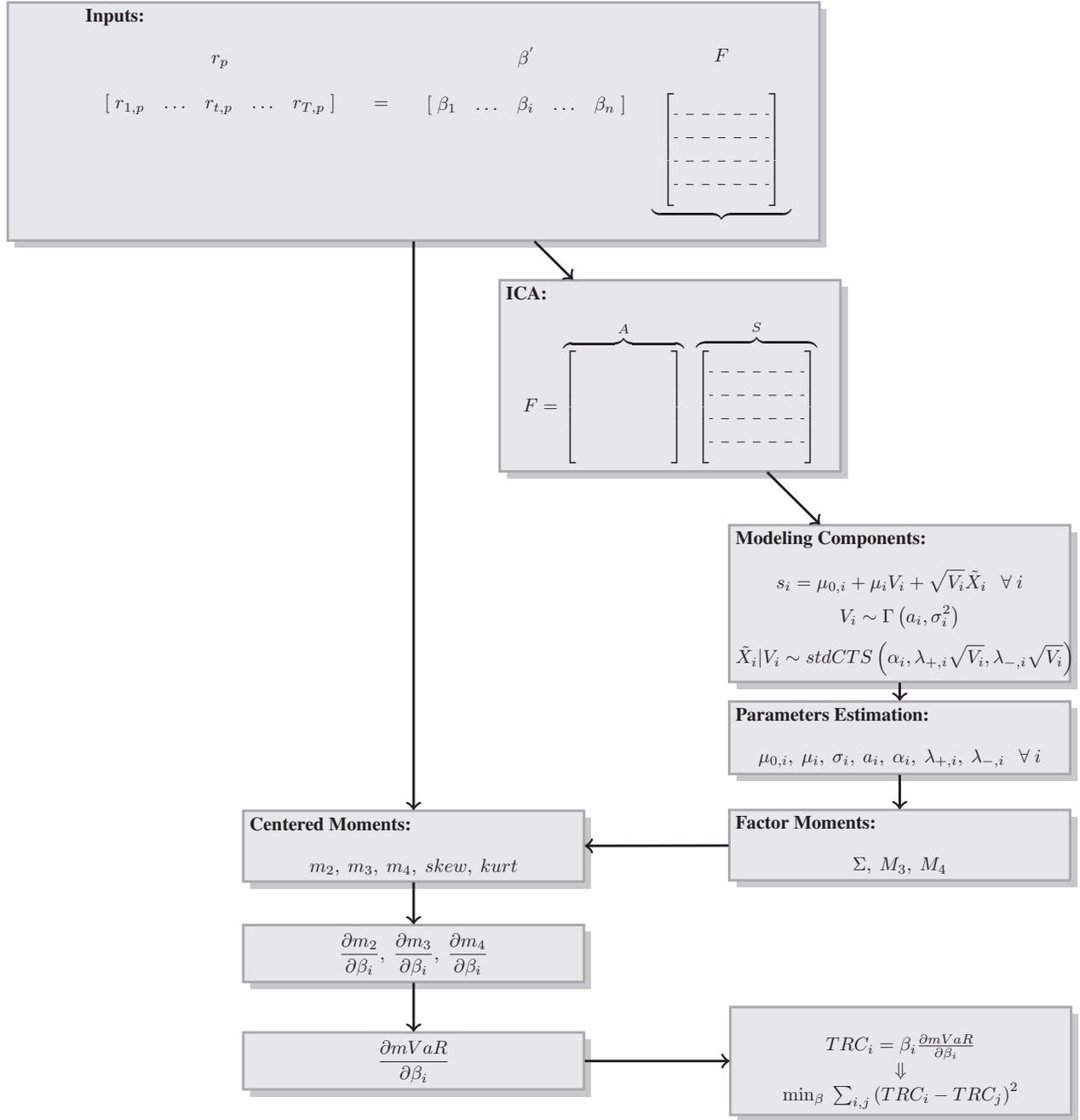}
\caption{Here we describe the main steps required in parametric risk parity portfolio construction. We start with a linear model for modeling portfolio returns and apply the ICA algorithm on the factor matrix. Each source signal $s_{i}$ is then modeled using the MixedTS distribution. The fitted parameters on the time series of each $s_{i}$  are used for the computation of the moments. The marginal risk contribution formulas (here we consider the modified VaR) require the partial derivatives of the centered moments that in our setup can be computed due to the independence assumption for the source signals. The last step for the portfolio construction is the optimization. \label{Fig:TIKZ}}
\end{figure}

We focus on homogeneous continuously differentiable risk measures for which risk contribution can be determined using the Euler's theorem for homogeneous functions \citep[see][for more details]{Tasche99}.\newline
Let $R(r)$ be an positive homogeneous risk measure, applying the Euler's theorem we get:
\begin{equation}
R(r)=\sum_{i=1}^{n} \beta_{i}\frac{\partial R(r)}{\partial \beta_{i}}=\sum_{i=1}^{n} TRC_{i}
\end{equation}where the Total Risk Contribution of the i-th risk factor is \citep[see][]{Tasche99} defined in equation \eqref{totalriskcontrib}. In particular the $TRC_{i}$ for the risk measures considered in this paper are listed below.
\begin{itemize}
\item For volatility:
\begin{equation}
TRC_{i}= \frac{\left(\Sigma \beta \right)_{i}}{\sqrt{\beta' \Sigma \beta}}
\label{Vol:Contr}
\end{equation}
where $\Sigma$ is the variance-covariance matrix of the factors.
\item For the value-at-risk \citep[see][for a complete treatment]{Gourieoux2000}:

\begin{equation}
TRC_{i}=-E\left[F_{i}\left|r=VaR_{\alpha}(r)\right.\right] \beta_{i}
\label{VaR:cont}
\end{equation}where $VaR_{\alpha}(r)$ is the value-at-risk of the portfolio evaluated at the level $\alpha$.
\item For the expected shortfall \citep[see][for more details]{Tasche02}:
\begin{equation}
TRC_{i}=-E\left[F_{i}\left| r \leq -VaR_{\alpha}(r) \right.\right] \beta_{i}
\label{ES:cont}
\end{equation}
\end{itemize}
The total risk contribution for a given factor can be easily computed using the historical approach. Indeed, we need only the matrix containing in the first column the vector $r$ while in the other columns 
we put the factor returns. Consider for example to the Value at Risk that is the quantile of a distribution. We take the complete data matrix and order all the data following the column of portfolio returns. Observe that once sorted the matrix we have all the information needed for risk decomposition. The marginal contribution to risk for the factors is then computed on the sorted factor columns. However as observed in \cite{boudt2007estimation} the estimating results obtained using historical Value-at-Risk and historical Expected Shortfall, have a large variation in the out-of-sample observations than those based on a correctly specified parametric class of distributions.\newline
In a non-gaussian parametric framework, the modified VaR proposed in \cite{Zangari1996} and the modified ES developed in \cite{boudt2007estimation} seem to be attractive approaches since both measures preserve the homogeneity property and they can be easily computed once the multivariate moments of the factors are available.
Using \eqref{Fact:Mod}, we model each asset return as a weighted average of factor returns. The mean vector for the factors is $\mu$ while $\Sigma$ is their variance-covariance matrix of dimension $N \times N$.
Co-skewness factor matrix of dimension $N \times N^2$ is:
\begin{equation}
M_{3}=E[(F-\mu)(F-\mu)'\otimes(F-\mu)']
\end{equation}while their co-kurtosis matrix is of dimension $N \times N^3$:
\begin{equation}
M_{4}=E[(F-\mu)(F-\mu)'\otimes(F-\mu)'\otimes(F-\mu)']
\end{equation} where $\otimes$ denotes the Kronecker product. The second, third and fourth order centered moments of the vector $r$ are respectively:
 \begin{equation}
\left\{ 
\begin{array}{l}
m_{2}=\beta' \Sigma \beta\\
m_{3}=\beta' M_{3} (\beta \otimes \beta) \\
m_{4}= \beta' M_{4} (\beta \otimes \beta\otimes \beta)\\

\end{array}
\right.
\end{equation}The skewness (skew)and kurtosis (kurt) are defined based on the centered moments :
\begin{equation}
skew=\frac{m_{3}}{m_{2}^{\frac{2}{3}}}
\end{equation}and
\begin{equation}
kurt=\frac{m_{4}}{m_{2}^{2}}-3
\end{equation}In order to compute $\Sigma$, $M_{3}$ and $M_{4}$ and consequently the centered moments, we need the multivariate distribution for the factor returns or their dependence structure by means of a copula function. Here we face the problem from a different point of view, that is we look for the underlying independent factors that generate the observed returns. In practice, the ICA analysis \citep[see][]{Hyvarinen99} applied to the factors simplifies the computation of $\Sigma$, $M_{3}$ and $M_{4}$ since:
\begin{equation}
F=AS
\label{ICA decomp}
\end{equation}where in $S=[S_{1}.....S_{N}]'$ we have the original sources  and $A$ is the mixing matrix to be estimated. Each signal is modeled using the MixedTS, i.e:
\begin{equation}
S^i \sim \mu_{0}^i+ \mu^i V^i + \sqrt{V^i}\tilde{X}^i
\end{equation}As shown in Appendix \ref{Apx:2} 
the computation of the elements of the moment matrices is quite easy and fast due to the factor independence:
\begin{equation}
\left\{ 
\begin{array}{l}
\bigskip
\Sigma^{ik}_{2}=\sum_{j=1}^{N}a_{ij}a_{kj} \sigma^2(s_{j})\\ \bigskip
M^{ikl}_{3}=\sum_{j=1}^{N}a_{ij}a_{kj}a_{lj} skew(s_{j})\\ 
M^{iklm}_{4}=\sum_{j=1}^{N}a_{ij}a_{kj}a_{lj}a_{mj}kurt(s_{j})\\
\end{array}
\right.
\label{Comoments:ICA}
\end{equation}Computed the moments and co-moments, the modified VaR is obtained using the formula derived in \cite{Zangari1996}:
\begin{equation}
mVaR_{\alpha}(r)=-\beta'\mu-\sqrt{m_{2}}\Phi^{-1}(\alpha)+\sqrt{m_{2}}C(z_{\alpha},skew,kurt)
\label{modVaR} 
\end{equation} where the quantity:
\begin{equation}
C(z_{\alpha},skew,kurt)= \left[-\frac{1}{6}(z^2_{\alpha}-1)skew-\frac{1}{24}(z^3_{\alpha}-3z_{\alpha})kurt+\frac{1}{36}(2z^3_{\alpha}-5z_{\alpha})skew^2 \right]
\end{equation}corrects the Gaussian VaR by considering skewness ($skew$) and kurtosis ($kurt$) of the return vector $r$ and $z_{\alpha}=\Phi^{-1}(\alpha)$. Observe that $\Phi()$ denotes the distribution of the standard normal while its inverse is used for the quantile determination.
Modified Expected Shortfall defined in \cite{boudt2007estimation} is a linear transformation of the expected value of the returns below the $\alpha-$Cornish fisher quantile where the second order Edgeworth expansion of the true distribution is considered:
\begin{equation}
mES_{\alpha}(r)=-\beta'\mu-\sqrt{m_{2}}E_{G_{2}}[z\left|z\leq g_{\alpha}\right.]
\label{modES}
\end{equation}
with $g_{\alpha}=G^{-1}_{2}(\alpha)$.
The extended formula is:

\begin{eqnarray}
E_{G_{2}}\left[z\left|z\leq g_{\alpha}\right.\right] &=& 
-\frac{1}{\alpha}\left\{\phi(g_{\alpha})+\frac{1}{24}\left[I^{4}-6I^{2}+3\phi(g_{\alpha})\right]kurt+\frac{1}{6}\left[I^{3}-3I\right]skew\right.\\
&+&\left.\frac{1}{72}\left[I^{6}-15I^{4}+45I^{2}-15\phi(g_{\alpha})\right]skew^{2} \right\}
\end{eqnarray}
where 

\begin{equation}
I^{q}=\left\{ 
\begin{array}{l}
\prod^{q/2}_{j=1}(\frac{\prod^{q/2}_{j=1} 2j}{\prod^{i}_{j=1} 2j})g^{2i}_{\alpha}\phi(g_{\alpha})+(\prod^{q/2}_{j=1} 2j)\phi(g_{\alpha}) \ \ \ \ \ \ \ \ \ \ \ \ \ \ \ \ \ \ \ \ \ \ \ \ \ \ \ for \ \ q \ \ even\\
\prod^{q^{*}}_{j=0}(\frac{\prod^{q^{*}}_{j=0} (2j+1)}{\prod^{i}_{j=0} (2j+1)})g^{2i+1}_{\alpha}\phi(g_{\alpha})-(\prod^{q^{*}}_{j=0} (2j+1))\phi(g_{\alpha}) \  \ \  \ \ \ \ \ \ \ \ \ \ for \ \ q \ \ odd \\

\end{array}
\right.
\end{equation} and $q^{*}=\frac{q-1}{2}$. The partial derivatives formulas for the centered moments are:
 \begin{equation}
\left\{ 
\begin{array}{l}
\frac {\partial m_{2}}{\partial \beta_{i}}=2\left(\Sigma \beta\right)_{i}\\
\frac {\partial m_{3}}{\partial \beta_{i}}=3\left[M_{3} (\beta \otimes \beta)\right]_{i} \\
\frac {\partial m_{4}}{\partial \beta_{i}}= 4\left[ M_{4} (\beta \otimes \beta\otimes \beta)\right]_{i}\\

\end{array}
\right.
\end{equation}Modeling the source signals with the Mixed Tempered Stable  makes the computations easier. Partial derivatives allow us to obtain the total risk contribution for factors for modified VaR using the following formula:


\begin{eqnarray*}
\frac{\partial mVaR_{\alpha}(r)}{\partial \beta_{i}} &=& -\mu_{i}- \frac{\partial m_{2}}{\partial \beta_{i}} \frac{1}{2\sqrt{m_{2}}}\Phi^{-1}(\alpha)\\
&+&\frac{\partial m_{2}}{\partial \beta_{i}} \frac{1}{\sqrt{m_{2}}} \left[-\frac{1}{12}(z^2_{\alpha}-1)skew-\frac{1}{48}(z^3_{\alpha}-3z_{\alpha})kurt+\frac{1}{72}(2z^3_{\alpha}-5z_{\alpha})skew^2 \right]\\
 &+& \sqrt{m_{2}}\left[-\frac{1}{6}(z^2_{\alpha}-1)\frac{\partial skew}{\partial \beta_{i}}-\frac{1}{24}(z^3_{\alpha}-3z_{\alpha})\frac{\partial kurt}{\partial \beta_{i}}+\frac{1}{18}(2z^3_{\alpha}-5z_{\alpha})skew\frac{\partial skew}{\partial \beta_{i}} \right]
\end{eqnarray*}In the same way,  total risk contributions for modified Expected Shortfall can be obtained using a similar formula given in \cite{boudt2007estimation}. The derivative of \eqref{modES} requires straightforward calculations but can be implemented directly using standard algebra in any programming language. \newline
In Figure \ref{Fig:TIKZ} we give a detailed description of the entire procedure described in this Section.
\bigskip

Insert here Figure \ref{Fig:TIKZ}.

\section{Empirical analysis}
In this Section we show step-by step how to obtain a risk parity portfolio using the MixedTS for modeling the source signals in the market. The dataset is composed by daily log returns of the Vanguard Fund Index (VFIAX) which replicates the performance of the S$\&$P 500 and the  the ten sector indexes: Utility, Telecommunications, Materials, Information Technology, Industrial, Health, Financial, Energy, Consumption Staple and Consumption Discretionary that are  considered as risk factors. The dataset refers to the period from 24/06/2010  to 10/07/2013. In Table \ref{tab:MainStat} we give the main statistics of the time series we use in this Section. Observe that they result to be negatively skewed and with tails heavier than what can be predicted from the normal distribution. The higher volatility of the Financial sector reflects the crisis that in this time frame was at its ultimate phase. 
\bigskip

Insert here Table \ref{tab:MainStat}.

\begin{figure}[H]
	\centering
		\includegraphics[width=1\textwidth]{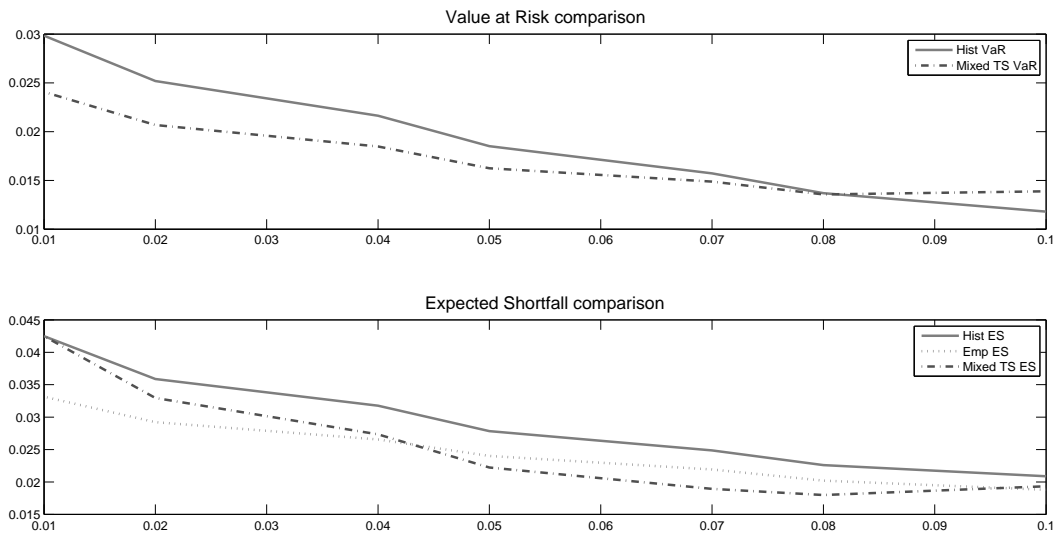}	
	\caption{In the figure we show the Value at Risk of the VFIAX index computed for the period  24/06/2010 - 10/07/2013 for $\alpha \in (0.01:0.1)$ using both the historical approach and the inversion formula for the MixedTS characteristic function. For the Expected Shortfall we perform the same analysis but in addition we give the empirical (robust) ES since it does not consider in the mean the data lower than an $\alpha_{1}$- quantile that is we do not into account extreme values.\label{fig:ConfrontoMisureDiRischio22_04_Con_Emp_ES}}
\end{figure}

\bigskip
As a first step we want to show the univariate risk measures we get when the MixedTS distribution is used directly for modeling the observed time series. We fit the MixedTS distribution to the returns of the VFIAX  fund and compare the historical VaR and ES for the entire period with the respective parametric versions using formulas \eqref{modVaR} and \eqref{modES}. The analysis is done for confidence levels $\alpha$ ranging  in $(0.01; 0.1)$. The historical VaR for $\alpha<0.08$ results to be lower than the VaR computed using the MixedTS as observed in Figure \ref{fig:ConfrontoMisureDiRischio22_04_Con_Emp_ES}. The difference becomes noticeable only for $\alpha < 0.05$. The results concerning the ES highlight more the importance of choosing parametric or non-parametric methods for measuring risk. In fact, we have that the historical method for the ES gives higher values than the MixedTS based ES. The difference is bigger for  $\alpha \in (0.04: 0.08)$.  Notice that ES is a conditional mean and is highly influenced by extreme values. We consider the comparison with the empirical (robust see \cite{Cont2010}) ES that is a trimmed mean since its definition for $0 < \alpha_{1} < \alpha_{2} < 1$ is:
\begin{equation}
ES^{Empi}_{\alpha}(Y)= \frac{1}{(\alpha_{2}-\alpha_{1})}\int^{\alpha_{2}}_{\alpha_{1}} VaR_{u}(F_{Y}) du
\end{equation}The robust ES is less sensible to extreme values and the empirical quantities we get are similar to the MixedTS based ES.
\bigskip

Insert Figure \ref{fig:ConfrontoMisureDiRischio22_04_Con_Emp_ES}.

\bigskip
In the next step we consider the returns of the VFIAX fund as a linear combination of the sector returns. As described in the previous section, we perform an ICA analysis on the matrix whose rows are the sector indexes returns. The output of this algorithm is the mixing matrix in Table \ref{Mix:Matr} and the time series of the underlying signals. Following the idea of the algorithm, each market return time series is a linear transformation of the independent factors that lead the market.
\bigskip

Insert here Table \ref{Mix:Matr}. 

\bigskip
We fit the MixedTS to the independent factor time series. The fitted parameters refered to the first window are reported in Table \ref{Comp:MixedTS}. Particular attention deserves the parameter $\alpha$ since for $\alpha=2$ we get the Variance Gamma distribution. We notice that only the fourth and the fifth components can be modeled with the Variance Gamma. The first four moments of each component are computed once we have the parameters. As discussed before, the independence hypothesis in the ICA algorithm gives rise to the analytic higher order moments for the matrix of the portfolio factors, i.e we compute the moments of the matrix whose rows are the returns of each sector.
\bigskip

Insert here Table \ref{Comp:MixedTS}.

\bigskip

\begin{figure}[H]
	\centering
		\includegraphics[width=1\textwidth]{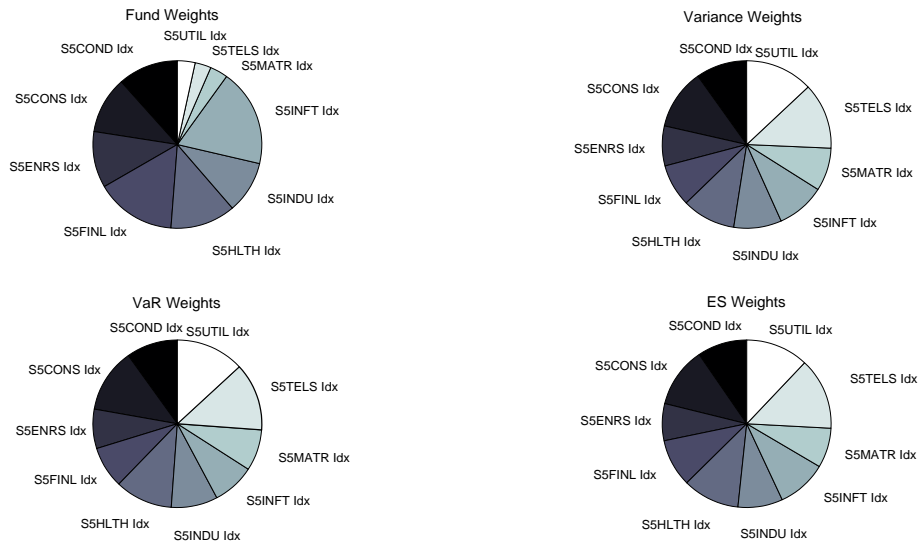}
	
	\caption{The four pies show the portfolio composition respectively of the VFIAX fund and the three risk parity portfolios for the homogeneous risk measures: volatility, VaR and ES. The fund weights refer to the closing date 24/06/2011 and the risk parity portfolios are computed at the same date with 250 days ahead of data. Observe that the choice of the risk measure does not have a great effect on the weights given to each sector based on our analysis and considering the MixedTS distribution for modeling the source signals.\label{fig:FiguraPesiPrimoFinestrino}}
\end{figure}

Insert here Figure \ref{fig:FiguraPesiPrimoFinestrino}.

\bigskip

To have an intuition about our procedure we perform a rolling analysis and compare the out-of sample performances of the VFIAX fund with the three risk based portfolios. We consider the period  24/06/2011 till 10/07/2013 with 250 closing prices as in sample  data and the following 50 closing prices as out of sample data. In Table \ref{tab:Results} we report the mean of returns for the S\&P 500 index, the VFIAX Fund index, and for the risk parity parametric portfolios for three risk measures: Volatility, VaR and ES. in the  rolling window analysis.  First we give the results for each out of sample window and then the mean and standard deviation of all out of sample results. Observe that the choice of the risk measure does not have a great effect on the weights given to each sector based on our analysis and considering the MixedTS distribution for modeling the source signals.
\bigskip

Insert here Table \ref{tab:Results}.

\bigskip

\begin{figure}[H]
	\centering
		\includegraphics[width=1\textwidth]{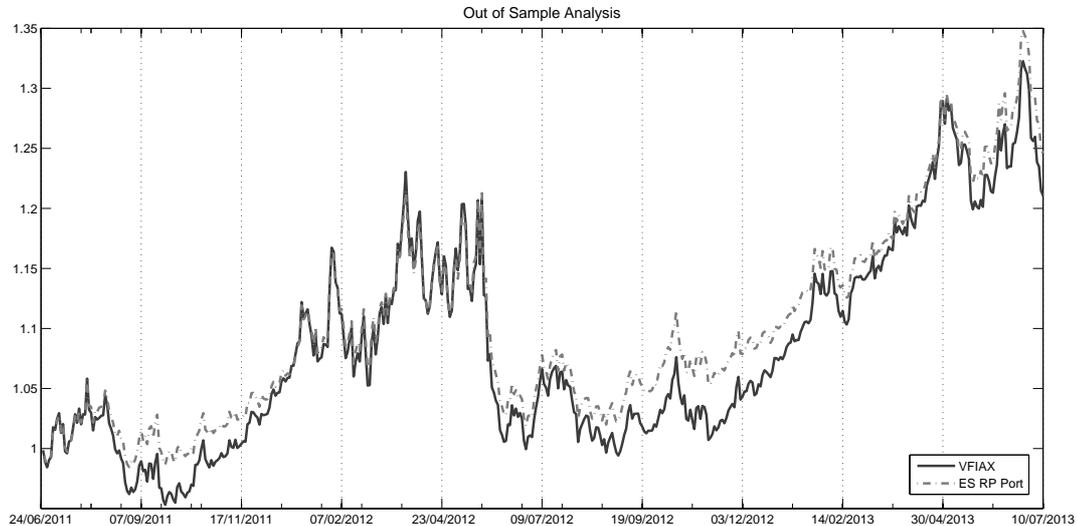}
	
	\footnotesize{\caption{In this Figure we plot the out of sample performance of two portfolios: the VFIAX fund and the risk parity portfolio when the risk measure considered is the Expected Shortfall. The analysis refers to the period 24/06/2011 till 10/07/2013 and we have rolling windows of 250 closing prices as in sample  data and the following 50 closing prices as out of sample data where we compare in terms of performance the strategy of the VFIAX fund and the parametric risk parity portfolio.\label{fig:OutOfSampleRiskParityNUOVA}}}
\end{figure}

In  Figure \ref{fig:OutOfSampleRiskParityNUOVA} we plot the out of sample performance of two portfolios: the VFIAX fund and the risk parity portfolio when the risk measure considered is the Expected Shortfall. From this plot we can immediately observe that the risk parity portfolio has better out-of sample performance. This result is valid for the other two risk parity based portfolios but we decided to show only one comparison since we think three similar plots would be redundant.
\bigskip

Insert here Figure \ref{fig:OutOfSampleRiskParityNUOVA}.

\bigskip
Then we decide to assess the statement that risk parity portfolios are well-diversified and consider the Gini index for measuring diversification. In fact the Gini index for equally weighted portfolios equals $0$ and $1$ when all the weight is given to one asset i.e for perfectly concentrated portfolios. In Table \ref{MisConc} we give the concentration measures for our portfolio based on the consistent estimator of the Gini index G:

\begin{equation}
G=\frac{1}{N-1}\left(N+1-2\left(\frac{\sum^{N}_{i=1} (N+1-i)y_{i}}{\sum^{N}_{i=1} y_{i}}\right)\right)
\label{Gini}
\end{equation}where the observations are ordered, i.e $y_{i}\leq y_{i+1}$.
\bigskip

Insert here Table\ref{MisConc}.

\bigskip


In particular, we report the respective indexes for each window of the rolling analysis. We find that risk parity portfolios based on the two risk measures Volatility and ES (its modified version) are less concentrated almost in all windows. The VFIAX fund weights follow the market capitalization of the sectors though the Gini index computed on these weights follows  the market. Risk parity portfolios based on VaR (modified version) seems to be more concentrated than the alternative optimized portfolios. In order to make an investment decision
we have to consider both performance and desired level of concentration. However, based on our results we have that risk parity portfolios are less concentrated and show better out-of sample performances than a passive strategy that can be for example investing on a fund as the VFIAX that replicates the S$\&$P500 returns.
\section{Conclusion}
In this paper we give the steps required in a parametric risk decomposition framework. The idea of applying the ICA analysis on the factors and modeling each source signal with the MixedTS distribution gives rise to the possibility of having analytical formulas for the moments and flexibility in capturing tail behavior.   
This approach can be applied to any setup that considers an homogeneous risk measure. In the paper we consider the Volatility, the VaR and ES being the three most used in the practice and in academia. Our results suggest that the decision of which risk measure to consider is not so relevant for the portfolio composition but we observe that the risk parity strategy generates well-diversified portfolios with good out-of sample performances.

\begin{appendices}
\section{Derivation of the moments}
\label{DerivMoment}
We derive the mean, variance, third and fourth order central moments of a MixedTS Random Variable.
A continuous random variable $Y$ is a Mixed Tempered Stable if it can be written as:
\begin{equation*}
Y=\mu_{0}+\mu V+ \sqrt{V} X
\end{equation*}
where $X$ given $V$ is a standardized tempered stable with parameters 
$stdCTS\left(\alpha,\lambda_{+}\sqrt{V},\lambda_{-}\sqrt{V}\right)$. 
We recall the formula for the cumulant of order n of the standardized tempered stable with parameters $\left(\alpha ,\lambda_{+} , \lambda_{-} \right)$:
\begin{equation*}
c_{n}\left(X\right)=\Gamma\left(n-\alpha \right)\left(\lambda_{+}^{\alpha-n}+\left(-1\right)^{n}\lambda_{-}^{\alpha-n}\right)C, \ \ \ n=2, \ldots 
\end{equation*} where the constant $C$ is fixed in order to ensure the standardization condition
\begin{equation*}
C=\frac{1}{\Gamma\left(2-\alpha\right)\left(\lambda_{+}^{\alpha-2}+\lambda_{-}^{\alpha-2}\right)}.
\end{equation*}In following we show how to determine the moments.\newline
\textbf{Mean}:
\begin{equation*}
E\left[Y\right]=\mu_{0}+\mu E\left[V\right].
\end{equation*}
\textbf{Variance}:
\begin{eqnarray*}
Var\left[Y\right]&=&E\left\{\left[\mu \left(V-E(V)\right)+ \sqrt{V} X\right]^{2}\right\}\\
&=&E\left\{\mu^2 \left(V-E(V)\right)^2+ V X^{2} +2 \mu \left(V-E(V)\right) \sqrt{V} X \right\}\\
\end{eqnarray*}Applying the linearity and the iteration properties of the expected value, we obtain:
\begin{eqnarray*}
Var\left[Y\right]&=&\mu^2 Var(V)+ E\left[V E\left(X^2\right. \left| V\right)\right]\\
&=&\mu^2 Var(V)+ E\left[V \right]\\
\end{eqnarray*}
\textbf{Third central moment}:
\begin{eqnarray*}
m_{3}\left(Y\right)&=&E\left\{\left[\mu \left(V-E(V)\right)+ \sqrt{V} X\right]^{3}\right\}\\
&=&E\left\{\left[\mu^3 \left(V-E(V)\right)^3+3\mu^2 \left(V-E(V)\right)^2 \sqrt{V} X+ 3\mu \left(V-E(V)\right) V X^2
+V^{3/2} X^3\right]\right\}
\end{eqnarray*}Applying the iteration property we show that:
\begin{equation*}
E\left[\sigma\mu^2 \left(V-E(V)\right)^2 \sqrt{V} X\right]=0
\end{equation*}then
\begin{equation*}
m_{3}\left(Y\right)=\mu^3m_{3}\left(V\right)+ 3\mu E\left[\left(V-E(V)\right) V X^2
\right]+E\left[ V^{3/2} X^3\right]
\end{equation*}Using:
\begin{equation*}
E\left[\left(V-E(V)\right) V X^2\right]=Var(V)
\end{equation*}and:
\begin{eqnarray*}
E\left[ V^{3/2} X^3\right]&=&E\left[V^{3/2}E\left(X^3\right.\left|V\right)\right]\\
&=&E\left[V^{3/2}\frac{\Gamma\left(3-\alpha \right)\left(\lambda_{+}^{\alpha-3}+\left(-1\right)^{3}\lambda_{-}^{\alpha-3}\right)}{\Gamma\left(2-\alpha\right)\left(\lambda_{+}^{\alpha-2}+\lambda_{-}^{\alpha-2}\right)}\frac{V^{\alpha/2-3/2}}{V^{\alpha/2-2/2}}\right]
\end{eqnarray*}By straightforward calculation and using the property of the Gamma function, we get:
\begin{equation*}
E\left[ V^{3/2} X^3\right]=\left(2-\alpha\right)\frac{\left(\lambda_{+}^{\alpha-3}-\lambda_{-}^{\alpha-3}\right)}{\left(\lambda_{+}^{\alpha-2}+\lambda_{-}^{\alpha-2}\right)}E\left[V\right]
\end{equation*}Finally the third central moment is:
\begin{equation*}
m_{3}\left(Y\right)=\mu^3m_{3}\left(V\right)+3\mu Var(V)+\left(2-\alpha\right)\frac{\left(\lambda_{+}^{\alpha-3}-\lambda_{-}^{\alpha-3}\right)}{\left(\lambda_{+}^{\alpha-2}+\lambda_{-}^{\alpha-2}\right)} E\left[V\right]
\end{equation*}
\textbf{Fourth central moment}:
\begin{eqnarray*}
m_{4}\left(Y\right)&=&E\left\{\left[\mu \left(V-E(V)\right)+ \sqrt{V} X\right]^{4}\right\}\\
&=& \mu^4 k\left(V\right)+4E\left\{\left[\mu^3\left(V-E(V)\right)^3 \sqrt{V} X\right]\right\}
+ 6E\left\{\left[\mu^2 \left(V-E(V)\right)^2 V X^2\right]\right\}\\
&+& 4\mu E\left\{\left[ \left(V-E(V)\right) V^{3/2} X^3\right]\right\}+E\left[ V^2 X^4\right]\\
\end{eqnarray*}We need to calculate explicitly only the last two terms since the others were determined before.
\begin{eqnarray*}
E\left[ \left(V-E(V)\right) V^{3/2} X^3\right]&=&\left(2-\alpha\right)\frac{\lambda_{+}^{\alpha-3}-\lambda_{-}^{\alpha-3}}{\lambda_{+}^{\alpha-2}+\lambda_{-}^{\alpha2}}Var(V)\\
E\left[V^2 X^4\right]&=&E\left[V^2E\left[X^{4}\left|V\right.\right]\right]\\
&=&E\left[V^2\frac{\Gamma\left(4-\alpha\right)}{\Gamma\left(2-\alpha\right)}\frac{\left(\lambda_{+}^{\alpha-4}+\lambda_{-}^{\alpha-4}\right)}{\left(\lambda_{+}^{\alpha-2}+\lambda_{-}^{\alpha-2}\right)}\frac{V^{\alpha/2-2}}{V^{\alpha/2-1}}\right]\\
&=&(3-\alpha)(2-\alpha)\frac{\left(\lambda_{+}^{\alpha-4}+\lambda_{-}^{\alpha-4}\right)}{\left(\lambda_{+}^{\alpha-2}+\lambda_{-}^{\alpha-2}\right)}E\left[V\right]\\
\end{eqnarray*}then
\begin{eqnarray*}
m_{4}\left(Y\right) &=& \mu^4 m_{4}(V)+6 \mu^2 E\left[ \left(V-E(V)\right)^2 V \right]+4 \mu \left(2-\alpha\right)\frac{\lambda_{+}^{\alpha-3}-\lambda_{-}^{\alpha-3}}{\lambda_{+}^{\alpha-2}+\lambda_{-}^{\alpha-2}}Var(V)\\
&+& (3-\alpha)(2-\alpha)\frac{\left(\lambda_{+}^{\alpha-4}+\lambda_{-}^{\alpha-4}\right)}{\left(\lambda_{+}^{\alpha-2}+\lambda_{-}^{\alpha-2}\right)}E\left[V\right]
\end{eqnarray*}

\section{Moments using ICA}
\label{Apx:2}
We derive the components of the variance-covariance $\Sigma$ matrix in \eqref{Comoments:ICA} 
\begin{eqnarray*}
\Sigma_{2}^{ik}&=&E\left[\left\{ r_{i}- E\left[r_{i} \right] \right\}\left\{ r_{k}- E\left[r_{k} \right] \right\} \right] \\&=&E\left[\left\{\sum^N_{j=1}a_{ij}\left(s_{j}-E[s_{j}]\right) \right\} \left\{\sum^N_{j=1}a_{kj}\left(s_{j}-E[s_{j}]\right) \right\}\right]\\&=&\sum_{j=1}^{N}a_{ij}a_{kj} \sigma^2(s_{j})\\
\end{eqnarray*}Let us now compute the element $M^{ikl}_{3}$ as:
\begin{eqnarray*}
M^{ikl}_{3}&=& E\left[ \left\{ r_{i}- E\left[r_{i} \right] \right\}\left\{ r_{k}- E\left[r_{k} \right] \right\}\left\{ r_{l}- E\left[r_{l} \right] \right\}\right]\\&=&E\left[\left\{\sum^N_{j=1}a_{ij}\left(s_{j}-E[s_{j}]\right) \right\} \left\{\sum^N_{j=1}a_{kj}\left(s_{j}-E[s_{j}]\right) \right\}\left\{\sum^N_{j=1}a_{lj}\left(s_{j}-E[s_{j}]\right) \right\}   \right]\\&=& \sum^N_{j=1}a_{ij}a_{kj}a_{lj} skew(s_{j})
\end{eqnarray*}Let us now compute the element $M^{iklm}_{4}$ as:

\begin{eqnarray*}
M^{iklm}_{4}&=& E\left[ \left\{ r_{i}- E\left[r_{i} \right] \right\}\left\{ r_{k}- E\left[r_{k} \right] \right\}\left\{ r_{l}- E\left[r_{l} \right] \right\} \left\{ r_{m}- E\left[r_{m} \right] \right\}\right]\\&=&\sum^N_{j=1}a_{ij}a_{kj}a_{lj}a_{mj}kurt(s_{j})
\end{eqnarray*}

\end{appendices}


\bibliographystyle{chicago}
\bibliography{BiblioThesis}

\begin{table}[htbp]
\footnotesize
  \centering
    \begin{tabular}{rrrrrrr}
    \hline
          & Mean  & Std   & Skewness & Kurtosis & Max   & Min \\
    \hline
    VFIAX  & 5.22E-04 & 0.0111 & -0.4990 & 7.4284 & 0.0463 & -0.0690 \\
    COND & 7.93E-04 & 0.0119 & -0.5873 & 6.4336 & 0.0472 & -0.0690 \\
    CONS & 5.67E-04 & 0.0076 & -0.4175 & 6.0214 & 0.0332 & -0.0390 \\
    ENRS & 4.77E-04 & 0.0145 & -0.4215 & 6.8501 & 0.0687 & -0.0864 \\
    FINL & 4.00E-04 & 0.0159 & -0.3977 & 7.9692 & 0.0789 & -0.1052 \\
    HLTH & 6.69E-04 & 0.0096 & -0.4605 & 6.7295 & 0.0456 & -0.0540 \\
    INDU & 5.08E-04 & 0.0129 & -0.4854 & 6.3092 & 0.0495 & -0.0711 \\
    INFT & 4.31E-04 & 0.0121 & -0.2512 & 5.2089 & 0.0445 & -0.0596 \\
    MATR & 3.73E-04 & 0.0147 & -0.3828 & 5.9989 & 0.0593 & -0.0756 \\
    TELS & 5.25E-04 & 0.0096 & -0.2754 & 5.5523 & 0.0426 & -0.0550 \\
    UTIL & 3.07E-04 & 0.0086 & -0.1836 & 7.2391 & 0.0414 & -0.0563 \\
    \hline
    \end{tabular}%
		 \caption{Main statistics of the VFIAX and of the Sector Indexes  for the period considered. Note that they are all negatively skewed and their tails are heavier than the tails predicted from the normal distribution. The returns of the Financial sector have a higher volatility that is confirmed from the broader interval defined from the minimum and the maximum observation. \label{tab:MainStat}}
\end{table}

\begin{table}[H]
  \centering
 
    \begin{tabular}{cccccccccc}
    \hline
    \multicolumn{10}{c}{Mixing Matrix}   \\
    \hline
		\footnotesize
    -0.0113 & 0.0024 & -0.0040 & 0.0029 & -0.0016 & -0.0009 & -0.0078 & -0.0033 & 0.0012 & -0.0022 \\
    -0.0066 & 0.0010 & -0.0007 & 0.0038 & -0.0039 & -0.0004 & -0.0029 & -0.0005 & 0.0015 & -0.0002 \\
    -0.0140 & 0.0007 & -0.0069 & 0.0008 & -0.0062 & 0.0013 & -0.0072 & -0.0023 & 0.0051 & 0.0023 \\
    -0.0173 & 0.0010 & -0.0091 & 0.0050 & -0.0030 & 0.0037 & -0.0034 & -0.0037 & 0.0039 & -0.0040 \\
    -0.0095 & 0.0011 & -0.0039 & 0.0032 & -0.0032 & -0.0004 & -0.0048 & 0.0019 & 0.0004 & -0.0010 \\
    -0.0117 & -0.0004 & -0.0053 & 0.0037 & -0.0038 & 0.0019 & -0.0085 & -0.0014 & 0.0040 & -0.0031 \\
    -0.0103 & 0.0032 & -0.0053 & 0.0024 & -0.0005 & -0.0024 & -0.0069 & -0.0009 & 0.0061 & -0.0017 \\
    -0.0128 & 0.0022 & -0.0074 & 0.0000 & -0.0068 & 0.0004 & -0.0078 & -0.0021 & 0.0046 & -0.0043 \\
    -0.0091 & -0.0036 & -0.0017 & 0.0022 & -0.0026 & -0.0026 & -0.0018 & -0.0012 & 0.0016 & -0.0011 \\
    -0.0095 & 0.0000 & 0.0017 & 0.0009 & -0.0025 & 0.0011 & -0.0019 & 0.0009 & 0.0014 & -0.0006 \\
    \hline
    \end{tabular}
		 \caption{In this Table we show the Mixing Matrix obtained when we apply the ICA algorithm on the matrix whose vectors are the historical time series of returns from 24/06/2010 to 24/06/2011 of the ten sector indexes of the S\&P500. \label{Mix:Matr}}
\end{table}%

\bigskip

\begin{table}[H]
\footnotesize
  \centering
  
    \begin{tabular}{lcccccccccc}
    \hline
       & \multicolumn{1}{c}{I} & \multicolumn{1}{c}{II} & \multicolumn{1}{c}{III} & \multicolumn{1}{c}{IV} & \multicolumn{1}{c}{V} & \multicolumn{1}{c}{VI} & \multicolumn{1}{c}{VII} & \multicolumn{1}{c}{VIII} & \multicolumn{1}{c}{IX} & \multicolumn{1}{c}{X} \\
    \hline
    $\mu_0$ & 0.0989 & 0.1915 & 1.0361 & -0.0555 & 0.4227 & 0.5418 & 0.9911 & 0.7190 & 0.3449 & 0.7476 \\
    $\mu$    & -0.0719 & -0.0745 & -0.3914 & 0.0579 & -0.0674 & -0.0991 & -0.1763 & -0.1094 & -0.0688 & -0.1386 \\
    $\sigma$ & 0.6847 & 0.5991 & 0.5766 & 0.5132 & 0.3285 & 0.4095 & 0.3798 & 0.3729 & 0.4490 & 0.4705 \\
    $a$     & 2.1983 & 2.5824 & 2.6360 & 3.8144 & 6.6537 & 6.0530 & 5.8454 & 6.3537 & 5.0876 & 5.0049 \\
    $\alpha$ & 0.8740 & 1.7955 & 0.6383 & 2.0000 & 1.9904 & 0.0594 & 0.0100 & 1.5698 & 0.0100 & 0.1282 \\
    $\lambda_+$ & 1.1631 & 1.3175 & 1.2307 & 1.2924 & 1.2891 & 1.5148 & 1.9890 & 1.6767 & 1.6033 & 1.8090 \\
    $\lambda_-$ & 1.2186 & 1.4375 & 2.1308 & 2.9084 & 2.9103 & 2.6869 & 2.4690 & 4.0004 & 2.5576 & 2.4291 \\
    \hline
    \end{tabular}%
		\caption{In this Table we report the fitted parameters of the MixedTS distribution  to the independent components obtained by applying the ICA algorithm the matrix containing the returns from 24/06/2010 to 24/06/2011 of the ten sector indexes of the S\&P500. The valies of the parameters $\alpha$ are important in order to have an immediate idea of which special case of the MixedTS can better fit the return time series. \label{Comp:MixedTS}}
\end{table}

\begin{table}[htbp]
\footnotesize
  \centering
  
    \begin{tabular}{rrrrrr}
    \hline
    
		\multicolumn{6}{c}{Out-of-sample results for each window} \\
    \hline
		w     & mean SPX & mean VFIAX & mean $RP_{Volatility}$ & mean $RP_{VaR}$ & mean $RP_{ES}$ \\
    \hline
		1     & -0.0213\% & -0.0209\% & 0.0278\% & 0.0312\% & 0.0302\% \\
    2     & 0.0293\% & 0.0311\% & 0.0189\% & 0.0208\% & 0.0200\% \\
    3     & 0.2045\% & 0.2058\% & 0.1654\% & 0.1631\% & 0.1698\% \\
    4     & 0.0290\% & 0.0289\% & 0.0235\% & 0.0229\% & 0.0231\% \\
    5     & -0.1132\% & -0.1102\% & -0.0876\% & -0.0895\% & -0.0934\% \\
    6     & -0.0920\% & -0.0867\% & -0.0442\% & -0.0455\% & -0.0491\% \\
    7     & 0.0481\% & 0.0466\% & 0.0503\% & 0.0502\% & 0.0509\% \\
    8     & 0.1327\% & 0.1315\% & 0.1015\% & 0.1008\% & 0.1034\% \\
    9     & 0.2913\% & 0.2940\% & 0.2467\% & 0.2473\% & 0.2564\% \\
    10    & -0.1267\% & -0.1275\% & -0.0672\% & -0.0672\% & -0.0719\% \\
    \hline
		\multicolumn{6}{c}{Global out-of-sample results} \\
    \hline
          & SPX   & VFIAX   & $RP_{Volatility}$ & $RP_{VaR}$   & $RP_{ES}$ \\
    \hline
		mean  & 0.0382\% & 0.0393\% & 0.0435\% & 0.0434\% & 0.0439\% \\
    std   & 0.01242 & 0.01241 & 0.01090 & 0.010862 & 0.011040 \\
    \hline
    \end{tabular}%

	\footnotesize{\caption{In this Table we give the mean of returns of the S\&P 500 index, VFIAX Fund index, risk parity parametric portfolios for three risk measures: volatility, VaR and ES. for the rolling windows analysis in the period  24/06/2011 till 10/07/2013 with 250 closing prices as in sample  data and the following 50 closing prices as out of sample data. First we give the results for each out of sample window and then the mean and standard deviations of all out of sample results. \label{tab:Results}}}
\end{table}

\begin{table}[htbp]
  \centering
 \footnotesize
    \begin{tabular}{ccccc}
    \hline
    w   & $G^{VFIAX}$ & $G^{Vol_{RP}}$ & $G^{VaR_{RP}}$ &$G^{ES_{RP}}$\\
    \hline
    1     & 0.301 & 0.194 & 0.247 & 0.197 \\
    2     & 0.301 & 0.166 & 0.248 & 0.235 \\
    3     & 0.302 & 0.178 & 0.222 & 0.189 \\
    4     & 0.301 & 0.194 & 0.247 & 0.197 \\
    5     & 0.300 & 0.198 & 0.244 & 0.185 \\
    6     & 0.297 & 0.200 & 0.231 & 0.198 \\
    7     & 0.297 & 0.186 & 0.218 & 0.203 \\
    8     & 0.294 & 0.181 & 0.206 & 0.177 \\
    9     & 0.299 & 0.193 & 0.246 & 0.150 \\
    10    & 0.301 & 0.179 & 0.233 & 0.187 \\
    \hline
    \end{tabular}%
		\footnotesize{\caption{In the Table we report the Gini index computed for each rolling window, in the period  24/06/2011 till 10/07/2013 with 250 closing prices as in sample  data and the following 50 closing prices as out of sample data, for the VFIAX fund and for the three risk parity portfolios based respectively on the homogeneous risk measures: Volatility, VaR and ES. Following the risk parity approach we get less concentrated portfolios especially in the Volatily and mES based risk parity portfolios. \label{MisConc}}}
\end{table}%

\end{document}